\documentclass[a4paper,11pt]{article}

\usepackage{jcappub} 
\usepackage[normalem]{ulem}
\usepackage{xcolor}
\usepackage{physics}
\usepackage{slashed}
\usepackage{caption}
\usepackage{subfig}
\usepackage{siunitx}
\usepackage{comment}

\usepackage[dvipsnames,table,xcdraw]{xcolor}

\usepackage[T1]{fontenc} 

\newcommand{\Lag}{\mathcal{L}}

\title{How invisible can QCD axions be? 
From Supernova emission to Cherenkov signals}

\author[a]{Luca Di Luzio,}
\author[b,c]{Vincenzo Fiorentino,}
\author[d,e]{Maurizio Giannotti,}
\author[a,f]{Alessandro Lella,}
\author[b,g]{and Federico Mescia}

\affiliation[a]{Istituto Nazionale di Fisica Nucleare, Sezione di Padova, Via F.~Marzolo 8, 35131 Padova, Italy}
\affiliation[b]{Istituto Nazionale di Fisica Nucleare, Laboratori Nazionali di Frascati, Via E. Fermi 54, 00044 Frascati, Italy}
\affiliation[c]{
Dipartimento di Matematica e Fisica, Universit\`a Roma Tre, Via della Vasca Navale 84, I-00146 Rome, Italy}
\affiliation[d]{Centro de Astrof\'{i}sica y F\'{i}sica de Altas Energ\'{i}as (CAPA), University of Zaragoza, Zaragoza 50009, Arag\'{o}n, Spain}
\affiliation[e]{Department of Chemistry and Physics, Barry University, Miami Shores, FL 33161, United States of America}
\affiliation[f]{Dipartimento di Fisica e Astronomia ``Galileo Galilei'', Universit\`a di Padova, Via F.~Marzolo 8, 35131 Padova, Italy, Italy}
\affiliation[g]{On leave of absence from Universitat de Barcelona}

\emailAdd{luca.diluzio@pd.infn.it}
\emailAdd{vincenzo.fiorentino@lnf.infn.it}
\emailAdd{mgiannotti@barry.edu}
\emailAdd{alessandro.lella@unipd.it}
\emailAdd{federico.mescia@lnf.infn.it}

\abstract{
We investigate the scenario of \emph{maximally invisible axions},  namely QCD axions whose interactions with matter arise exclusively from the irreducible coupling  
to gluons responsible for solving the strong-CP problem.
We first analyze the production of such axions in core-collapse supernovae. In particular, we derive the corresponding SN 1987A cooling constraint and compute the emission spectra for the dominant production channels, namely nucleon-nucleon bremsstrahlung and pion conversion.
We then investigate the prospects for detecting maximally invisible axions in Cherenkov detectors, with the goal of establishing a robust lower bound on the overall detectability of QCD axions. 
}

\begin{document}
\maketitle
\flushbottom

\section{Introduction}
\label{sec:intro}
The strong CP problem stands as one of the most compelling puzzles in the
Standard Model (SM) of particle physics~\cite{tHooft:1976rip,Callan:1976je,Jackiw:1976pf}.
The QCD Lagrangian naturally accommodates a CP-violating term proportional to the
parameter $\bar{\theta}$, which receives contributions from both the QCD vacuum
angle and the phases of the quark mass matrix. Experimental constraints on the
neutron electric dipole moment (nEDM)~\cite{Abel:2020pzs} impose the stringent
bound $|\bar{\theta}| \lesssim 10^{-10}$, revealing a puzzling fine-tuning problem:
why should this dimensionless parameter, expected to be of order unity on
naturalness grounds, be so extraordinarily small?

The Peccei-Quinn (PQ) mechanism~\cite{Peccei:1977hh,Peccei:1977ur} provides an
elegant dynamical solution to this problem. By promoting $\bar{\theta}$ to a
dynamical field -- the axion~\cite{Weinberg:1977ma,Wilczek:1977pj} -- and
introducing a global $U(1)_{\rm PQ}$ symmetry that is anomalous under QCD, the
theory naturally relaxes to a CP-conserving vacuum. The axion, emerging as the
pseudo-Nambu-Goldstone boson of the spontaneously broken PQ symmetry, has since
become one of the most well-motivated candidates for physics beyond the SM. For
comprehensive reviews on axion physics, we refer the reader to
Refs.~\cite{Kim:2008hd,DiLuzio:2020wdo,Giannotti:2022euq}.

This compelling theoretical motivation has spurred an enormous experimental effort
to detect QCD axions. A rich and diverse program is currently underway, ranging
from haloscopes searching for axion dark
matter~\cite{ADMX:2018gho,HAYSTAC:2018rwy,CAPP:2020utb,McAllister:2017lkb,Ouellet:2019tlz,Gramolin:2020ict,QUAX:2020adt,DMRadio:2022pkf}
to helioscopes probing solar axion
production~\cite{CAST:2017uph,IAXO:2019mpb,IAXO:2024wss,IAXO:2020wwp}, and from
light-shining-through-wall experiments~\cite{ALPS:2009des,OSQAR:2015qdv,ALPSII:2025eri}
to searches exploiting astrophysical
observations~\cite{Giannotti:2017hny,Carenza:2024ehj,Caputo:2024oqc,DiLuzio:2021ysg}. A striking
feature of this landscape is that nearly all of these strategies rely on
\emph{model-dependent} axion couplings -- most notably the coupling to photons, or
the derivative couplings to nucleons and electrons -- whose magnitude is fixed not
by the solution of the strong CP problem itself, but by the details of the
ultraviolet (UV) completion of the axion model.

\noindent 
A defining feature of the QCD axion is its coupling to gluons,
\begin{equation}
\mathcal{L}_{a} \supset \frac{g_s^2}{32\pi^2} \frac{a}{f_a} G^{a}_{\mu\nu}
\tilde{G}^{a, \mu\nu} \,,
\label{eq:gluon_coupling}
\end{equation}
where $g_s$ is the strong coupling constant, $G^{a}_{\mu\nu}$ is the gluon field
strength tensor, $\tilde{G}^{a, \mu\nu}$ its dual, and $f_a$ is the axion decay
constant. This coupling is unavoidable: it generates the axion potential through
non-perturbative QCD effects and allows the dynamical relaxation of $\bar{\theta}$
to zero. 
Without it, the PQ mechanism would fail to solve the strong CP problem.
This is, in substance, the only \emph{essential} coupling of the QCD axion to SM
fields, gravitational interactions aside. By contrast, the couplings to photons and
to SM fermions are all model-dependent and can, in suitable UV completions, be significantly suppressed.

This observation motivates the central object of the present work. We study what we
term \emph{maximally invisible axions} -- QCD axions that couple to SM particles
only through the irreducible interactions mandated by the gluon coupling of
Eq.~\eqref{eq:gluon_coupling} (and by gravity). All other couplings, including those
to photons and fermions, are assumed to be suppressed. We emphasize that throughout
this work we consider exclusively QCD axions -- particles whose coupling to
gluons solves the strong CP problem -- and that this is the only model assumption we make. 
The maximally invisible scenario then represents the most pessimistic, yet
theoretically consistent, configuration from the standpoint of experimental
accessibility: the limit in which an axion that genuinely solves the strong CP
problem hides as effectively as the underlying physics allows.

Such a configuration is not merely an abstract construct. Explicit \emph{astrophobic}
axion models, in which the derivative couplings to nucleons and electrons are
simultaneously suppressed by means of generation-dependent PQ charges, have been
constructed and studied in
detail~\cite{DiLuzio:2017ogq,Bjorkeroth:2018ipq,Bjorkeroth:2019jtx,Badziak:2021apn,DiLuzio:2022tyc,Badziak:2023fsc,Takahashi:2023vhv,Badziak:2024szg,DiLuzio:2024vzg}. 
As we show in Appendix~\ref{app:model}, these constructions can be extended to
suppress the photon coupling as well, providing concrete existence proofs for the
maximally invisible axion scenario. 

Crucially, even maximally invisible axions are not entirely decoupled from SM
matter. The gluon coupling of Eq.~\eqref{eq:gluon_coupling} necessarily induces
interactions with nucleons and pions at low energies. Two irreducible couplings
emerge: the nEDM portal~\cite{Graham:2013gfa,Pospelov:1999mv,Crewther:1979pi},
which arises at one-loop level, and a shift-symmetry-breaking operator, involving nucleons and pions,
proportional
to the low-energy constant $\hat{c}_5$ in Heavy Baryon Chiral Perturbation Theory
(HBChPT)~\cite{Springmann:2024ret}. These couplings are fully determined by QCD and
independent of the UV completion, and therefore constitute a model-independent floor
for axion interactions with ordinary matter.

The phenomenological consequences of these irreducible couplings have only recently
begun to be explored.\footnote{We focus here on the relatively heavy [meV, eV] QCD axion mass window \cite{Cicoli:2026fqp}, which is of astrophysical interest. The light (sub-$\mu$eV) and ultra-light (sub-peV) mass windows 
can be probed instead by oscillatory phenomena induced by the 
model-independent coupling in Eq.~\eqref{eq:gluon_coupling}, under the additional hypothesis that the axion constitutes a non-negligible fraction of the dark matter. 
These include laboratory limits from 
EDM searches~\cite{Abel:2017rtm,Roussy:2020ily,JEDI:2022hxa,Schulthess:2022pbp}, 
radio-frequency atomic transitions~\cite{Zhang:2022ewz}, molecular clocks~\cite{Madge:2024aot}, 
and radioisotope decays \cite{Zhang:2023lem,Broggini:2024udi,Alda:2024xxa,Broggini:2026qxm}.} Reference~\cite{Lucente:2022vuo} treated the nEDM portal as a
model-independent feature of the QCD axion, revised the corresponding SN axion emission through this interaction, refined the SN 1987A cooling bound, and pointed out a potentially detectable high-energy photon signal from
$a+p\to p+\gamma$ in large water Cherenkov detectors such as Hyper-Kamiokande. 
More recently,
Ref.~\cite{Springmann:2024ret} identified a new, dominant production channel arising
from the $\hat{c}_5$ operator at tree level, which bypasses the one-loop suppression
of the nEDM operator.
This channel strengthens the model-independent bound on the
axion mass by approximately two orders of magnitude relative to the nEDM-portal
estimate, demonstrating that even maximally invisible axions leave non-trivial
astrophysical imprints.

These developments establish that the \emph{production} of maximally invisible
axions is by now relatively well understood, and they bring into focus a sharp,
complementary question that has not yet been addressed:
\emph{what are the prospects for directly detecting maximally invisible axions?}
Answering this question delineates the ultimate, model-independent reach of axion
searches. If maximally invisible axions can be detected for a given decay constant
$f_a$, then \emph{all} QCD axions with the same $f_a$, irrespective of their UV
completion, are in principle accessible to experiments. Conversely, if detection
proves impossible with current or foreseeable technology, there exists a class of
theoretically motivated
QCD axions that permanently evade direct experimental
scrutiny through their SM couplings.

In this work we provide a quantitative assessment of this question. Since the
maximally invisible axion has suppressed
photon coupling, experiments relying on
axion-photon conversion in magnetic fields, such as
helioscopes~\cite{IAXO:2019mpb}, are excluded by construction. The irreducible
$\hat{c}_5$ interaction instead manifests itself through pion production,
$a + N \to N + \pi^0$, so that the most natural -- and arguably the only viable
near-term -- detection handle is offered by large-volume water Cherenkov detectors
such as Hyper-Kamiokande~\cite{Hyper-Kamiokande:2018ofw}, where the subsequent decay
$\pi^0 \to \gamma\gamma$ produces a distinctive double-ring topology that
discriminates the signal from single-ring neutrino events~\cite{Cavan-Piton:2025nsj}.
This signature is conceptually distinct both from the oxygen de-excitation gamma
channel and from free-proton processes considered for derivatively-coupled axions,
and we focus on it throughout, leaving a systematic survey of alternative detectors
to future work.

Our analysis proceeds along the following line. 
We first compute the SN emission spectrum of maximally invisible axions, evaluating the $\hat{c}_5$-induced
nucleon bremsstrahlung and pion conversion channels and integrating over a
simulated SN profile (the \texttt{GARCHING} SFHo-s18.8 model~\cite{Garch}) --
to our knowledge, the first explicit presentation of this spectrum in the literature. 
In the course of this calculation we found it necessary to
revisit some expressions in the recent literature. Specifically, we report some differences with respect to the tree-level bremsstrahlung amplitudes of Ref.~\cite{Springmann:2024ret}.
With the spectrum in hand, we then estimate
the detection prospects in water Cherenkov detectors through the inverse
pion-conversion channel $a + N \to N + \pi^0$, taking Hyper-Kamiokande as a
benchmark and a nearby future Galactic SN as the source.

Our main finding is that, even under optimistic assumptions on source proximity
and detector performance, the expected signal lies well below one event for
values of $f_a$ consistent with the SN 1987A cooling bound. This conclusion,
however, turns out to be markedly sensitive to the modelling of the SN
environment.

This somewhat sobering result carries a clear physical message: QCD axions that
solve the strong CP problem may populate regions of parameter space where direct
detection through SM couplings is exceedingly difficult, even for values of $f_a$
regarded as accessible in benchmark scenarios. 
{In this regime, complementary
approaches -- cosmological probes, gravitational signatures, or the flavour-violating
signals that astrophobia necessarily entails -- become essential to explore the axion
sector.} We stress, however, that our quantitative conclusions are subject to known
astrophysical uncertainties, most notably the pion abundance in the SN
core~\cite{Fiorillo:2025gnd,Fore:2023gwv} and the modelling of the SN environment
itself; our negative result holds under the standard assumptions adopted throughout.

The remainder of this paper is organized as follows. In
Section~\ref{sec:couplings} we review the irreducible axion couplings induced by the
gluon interaction and establish the theoretical framework for our analysis.
Section~\ref{sec:sources} discusses the sources of maximally invisible axions, with
emphasis on supernova production, and presents the SN emission spectrum. In
Section~\ref{sec:detection} we analyze the Cherenkov detection strategy and compute
the expected signal. Section~\ref{sec:results} presents our main results and
discusses their implications. We conclude in Section~\ref{sec:conclusions} with a summary and
outlook. Appendix~\ref{app:model} provides an explicit model realization of
maximally invisible axions, and Appendix~\ref{sec:appB} collects the relevant
amplitudes and emission spectra.

\section{Irreducible axion couplings}
\label{sec:couplings}

In scenarios in which the direct axion-nucleon coupling in the PQ current $J^\mu_{\rm PQ}$ is absent, as is the case for astrophobic axion models, the standard leading axion production mechanism via one-pion exchange (OPE) is switched off. However, an \emph{irreducible} axion coupling remains.  

Let us consider two-flavour QCD coupled to a maximally invisible axion. The theory is described by the Lagrangian

\begin{equation}
\Lag_{\text{QCD}, a} =  -\frac{1}{4} G_{\mu \nu}^a G^{a, \mu \nu} + \bar{q}(i \slashed{D} - M_q) q +  \frac{1}{2} \partial_\mu a \partial^\mu a + \frac{g_s^2}{32 \pi^2} \frac{a}{f_a} G_{\mu \nu}^a \tilde{G}^{a, \mu \nu}.
\label{eq:b1}
\end{equation}

\noindent where $q = (u, d)^T$ and $M_q = \text{diag}(m_u, m_d)$. By performing a redefinition of the quark fields

\begin{equation}
q \to e^{\frac{ia}{2f_a} \gamma_5 Q_a}q,
\label{eq:b2}
\end{equation}

\noindent with $Q_a$ a flavour-space matrix satisfying $\Tr Q_a=1$, we can rotate away the axion-gluon coupling, obtaining in this new basis

\begin{equation}
\Lag_{\text{QCD}, a} = - \frac{1}{4} G_{\mu \nu}^{a} G^{a, \mu \nu} + \bar{q} (i \slashed{D}- M_a) q +  \frac{1}{2} \partial_\mu a \partial^\mu a,
\label{eq:b3}
\end{equation}

\noindent $M_a$ being the axion-dressed quark mass matrix:

\begin{equation}
M_a = e^{i \frac{a}{2f_a} Q_a} M_q e^{i \frac{a}{2f_a} Q_a}.
\label{eq:b4}
\end{equation}
We observe from Eq.~\eqref{eq:b3} that the axion-gluon interaction term introduces non-derivative axion couplings to the quark fields, which in turn give rise to similar couplings to nucleons and pions at low energies. In Ref.~\cite{Springmann:2024mjp}, these interaction terms were obtained at next-to-leading order in a HBChPT expansion:
\begin{equation}
\Lag_{\rm HBChPT} \supset \hat{c}_1 \bar{N} \ev{\chi_+} N + \hat{c}_5 \bar{N} \tilde{\chi}_+ N.
\label{eq:b5}
\end{equation}
\noindent Here $\chi_+ = u^\dagger \chi u^\dagger + u \chi^\dagger u$, with $\chi = 2 B M_a^\dagger$ and $u = \exp(i \pi^a \tau^a/ 2 f_\pi)$, and the parameter $B$ being related to the chiral condensate (see~\cite{Scherer:2002tk} for more details).

\begin{figure}[t]
\centering
\includegraphics[width=0.6\textwidth]{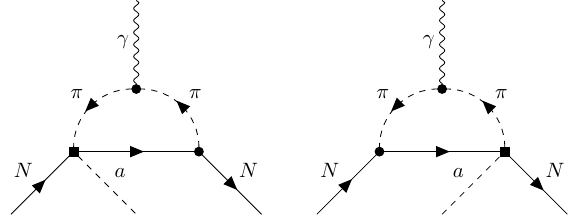}
\caption{One-loop diagrams generating the EDM operator through the $\hat{c}_5$ coupling. The square dot denotes the vertex arising from the $\hat{c}_5$ operator in Eq.~\eqref{eq:b6}.}
\label{fig:b1}
\end{figure}

While the $\hat{c}_1$ interaction term generates vertices with at least two axions, and therefore suppressed by a factor $1/f_a^2$, it was noted in~\cite{Springmann:2024ret} that the $\hat{c}_5$ term provides a model-independent axion coupling to pions and nucleons at leading order in $1/f_a$:

\begin{equation}
\Lag_{\rm HBChPT} \supset - \hat{c}_5 m_\pi^2 \frac{4z}{(1+z)^2} \bar{N} \qty(\frac{\pi^a a}{f_\pi f_a}) \tau^a N,
\label{eq:b6}
\end{equation}

\noindent where $z = m_u/m_d$, $m_\pi$ is the pion mass and $f_\pi$ is the pion decay constant.\footnote{Here, the standard choice $Q_a = M_q^{-1}/\Tr M_q^{-1}$ was made in order to avoid mass mixing between the axion and the $\pi^0$.} This coupling is irreducible, even in the maximally invisible scenarios. Moreover, it generates the nucleon EDM operator 

\begin{equation}
\Lag_a^{\text{EDM}} = -\frac{i}{2} \frac{C_{aN\gamma}}{m_N} \frac{a}{f_a} \bar{N} \gamma_5 \sigma_{\mu \nu} N F^{\mu \nu},
\label{eq:b7}
\end{equation}

\noindent 
with $C_{an\gamma} = - C_{ap\gamma} = 0.0033(15)$ \cite{Pospelov:1999mv}, arising from the 
one-loop diagrams in Fig.~\ref{fig:b1}. Thus, although slightly suppressed due to its isospin-violating nature, the operator in Eq.~\eqref{eq:b6} represents the dominant interaction with nucleons and pions for a maximally invisible axion.

\section{Core-collapse supernovae as sources of maximally invisible axions}
\label{sec:sources}

{The extreme temperatures and densities characterizing the core of massive stars undergoing gravitational collapse place core-collapse supernovae (SNe) among the most powerful factories for the production of feebly-interacting particles such as the QCD axion~\cite{Janka:2006fh, Mirizzi:2015eza, Carenza:2024ehj}. Indeed, typical temperatures of $30 - \SI{40}{MeV}$ and densities of the order of the nuclear saturation density $\rho \sim 10^{14} \,  \SI{}{g/cm^3}$ are expected in the SN core, thus allowing for a sizeable enhancement in the production of feebly-interacting particles. From this perspective, the possible future observation of a core-collapse SN would represent a unique opportunity to test axion physics.}

\begin{figure}[h]
\centering
\raisebox{0.22cm}{\includegraphics[width=.4\textwidth]{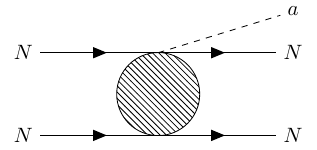}}
\qquad
\includegraphics[width=.3\textwidth]{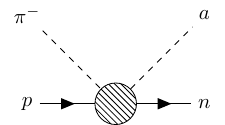}
\caption{Schematic representation of the Feynman diagrams for the two main axion production processes in core-collapse SNe: nucleon bremsstrahlung (\emph{left}) and pion conversion (\emph{right}).}
\label{fig:c1}
\end{figure}

Under typical SN conditions, the production of QCD axions in the nuclear medium of the core is expected to be dominated by nucleon-nucleon ($NN$) bremsstrahlung, schematically illustrated in the left panel of Fig.~\ref{fig:c1}. For benchmark axion models with derivative couplings to nucleons, the state-of-the-art calculation of the $NN$-bremsstrahlung axion emission rate has been carried out in Ref.~\cite{Carenza:2019pxu}. This analysis incorporates several important corrections to the traditional one-pion-exchange (OPE) approximation for the nuclear interaction potential, including contributions from two-pion exchange~\cite{Ericson:1988wr}, the finite mass of the exchanged pion~\cite{Stoica:2009zh}, in-medium modifications of the effective nucleon mass, and multiple-scattering effects~\cite{Raffelt:1991pw, Janka:1995ir}. Furthermore, Ref.~\cite{Springmann:2024mjp} introduced a framework to account for finite-density corrections to the axion-nucleon couplings themselves.

The contribution of the pion conversion process ({right panel of Fig.~\ref{fig:c1}}) to axion production in SNe was initially disregarded, because the fraction of thermal pions in the nuclear medium was believed to be extremely small, $Y_\pi \sim \mathcal{O}(10^{-4})$~\cite{Caputo:2024oqc}. However, in Ref.~\cite{Fore:2019wib} it was shown that the abundance of negatively charged pions can be increased by strong interactions, resulting in a pion conversion emission rate which is comparable or even dominant over bremsstrahlung~\cite{Carenza:2020cis}. The complete expression of the pion conversion emission rate, taking into account also the contact interaction term~\cite{Choi:2021ign}, as well as the role of the $\Delta (1232)$ resonance~\cite{Ho:2022oaw}, can be found in Refs.~\cite{Lella:2022uwi, Carenza:2023lci}. 
Nevertheless, some caution is warranted in this context. As emphasized in Ref.~\cite{Fiorillo:2025gnd}, the pion abundance in the supernova core remains uncertain and is currently the subject of active investigation (see, e.g., Ref.~\cite{Fore:2023gwv} for a recent analysis). This uncertainty can thus lead to potentially significant uncertainties in SN axion emissivities.
Under these assumptions, Ref.~\cite{Lella:2023bfb} ruled out derivative nucleon couplings $g_{aN}\gtrsim10^{-9}$ for $m_a\lesssim100\,$MeV, which would lead to an excessive shortening of the duration of the neutrino burst observed in coincidence to SN 1987A~\cite{Raffelt:1987yt}.

{Building upon previous studies considering axions with derivative nucleon couplings, in this work we evaluate for the first time 
 the axion emission rates from the SN nuclear medium in scenarios where derivative couplings vanish and nucleon couplings can arise solely from the $\hat{c}_5$ operator introduced in the previous section.} The relevant {expressions} for the Feynman amplitudes and the emission number spectra per unit time and volume are given in Appendix~\ref{sec:appB}. The emission spectrum is then obtained by integrating over the SN volume and over the emission time interval $\Delta t$

\begin{equation}
\dv{\mathcal{N}_a}{\omega_a} = \int_{0}^{\Delta t} \dd{t} \int_{\text{SN}} \dd[3]{\vec{r}} \dv{\dot{n}_a}{\omega_{a}}.
\label{eq:c1}
\end{equation}

\noindent As reference SN model, in our calculations we employ the 1D spherical-symmetric \texttt{GARCHING} group's SN model SFHo-s18.8 provided in Ref.~\cite{Garch} {(see, e.g. Refs.~\cite{Bollig:2020xdr,Caputo:2021rux,Caputo:2022mah,Lella:2022uwi,Lella:2023bfb,Lella:2024hfk,Lella:2024dmx,Springmann:2024ret,Springmann:2024mjp,Fiorillo:2025gnd} for some references emplying the same model)}, and we integrate over the entire duration of the SN cooling phase.

Being our goal the determination of the maximum observable axion flux in the maximally invisible scenario, we determined the value of $f_a$ saturating the SN 1987A cooling bound in our case. For this purpose, we first computed the axion emissivity associated to the two processes in Fig.~\ref{fig:c1} in the maximally invisible setup described in the previous section, as detailed in Appendix~\ref{sec:appB}. Then the axion luminosity is obtained as

\begin{equation}
L_a = 4 \pi \int_0^R \dd{r} Q_a(r),
\label{eq:c2}
\end{equation}

\noindent where $Q_a$ is the emissivity, see Eq.~\eqref{eq:B8}. $R=\SI{27}{km}$ is instead the radius of the SN core. The SN 1987A constraint can then be implemented exploiting the simple cooling criterion $L_a \lesssim L_\nu$, where $L_\nu \simeq 3 \times 10^{52}\, \SI{}{erg\, s^{-1}}$ is the observed neutrino luminosity in the SN 1987A event \cite{Raffelt:1987yt}. We obtain $L_a \times f_a^2 \simeq 3.96 \times 10^{62} \, \SI{}{erg^3 \, s^{-1}}$, yielding

\begin{equation}
f_a \gtrsim 7.17 \times 10^{7} \, \SI{}{GeV}.
\label{eq:c3}
\end{equation}
 
\begin{figure}[t]
\centering
\includegraphics[width=.6\textwidth]{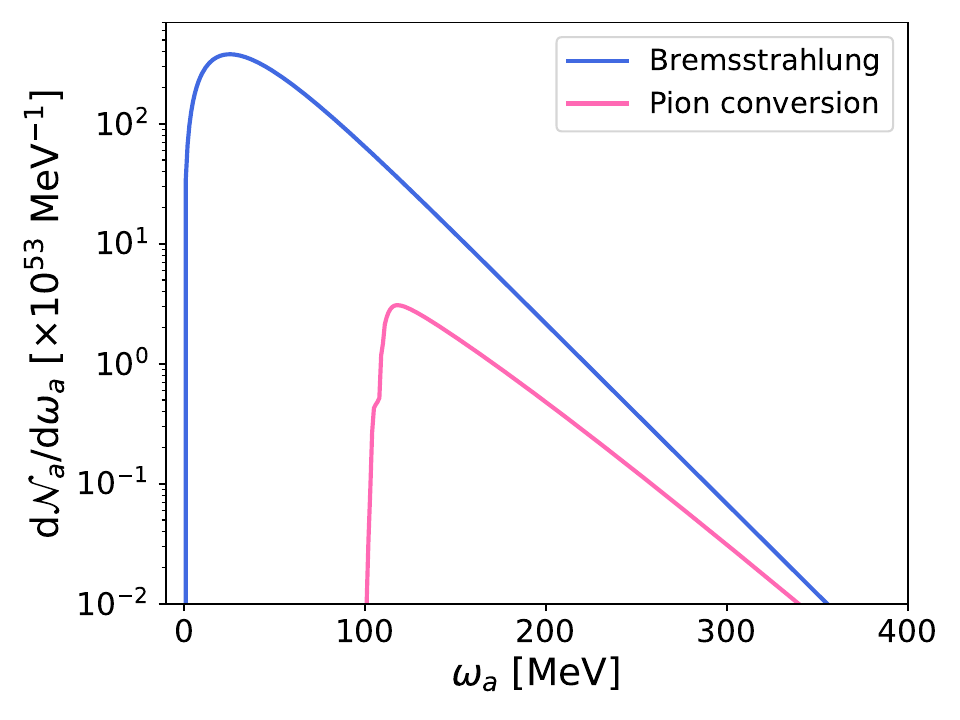}
\caption{Emission number spectra for maximally invisible axion production via nucleon bremsstrahlung (blue) and via pion conversion (pink) as functions of the axion energy, for $f_a$ saturating the corresponding SN 1987A cooling bound.}
\label{fig:c2}
\end{figure}

\noindent The results for the two production channels are shown in Fig.~\ref{fig:c2} for $f_a$ saturating the SN 1987A cooling limit. We observe that the dominant production process is nucleon bremsstrahlung, even above the threshold at $\omega_a = m_\pi$ where pion conversion is allowed.

\begin{figure}[t]
\centering
\includegraphics[width=.6\textwidth]{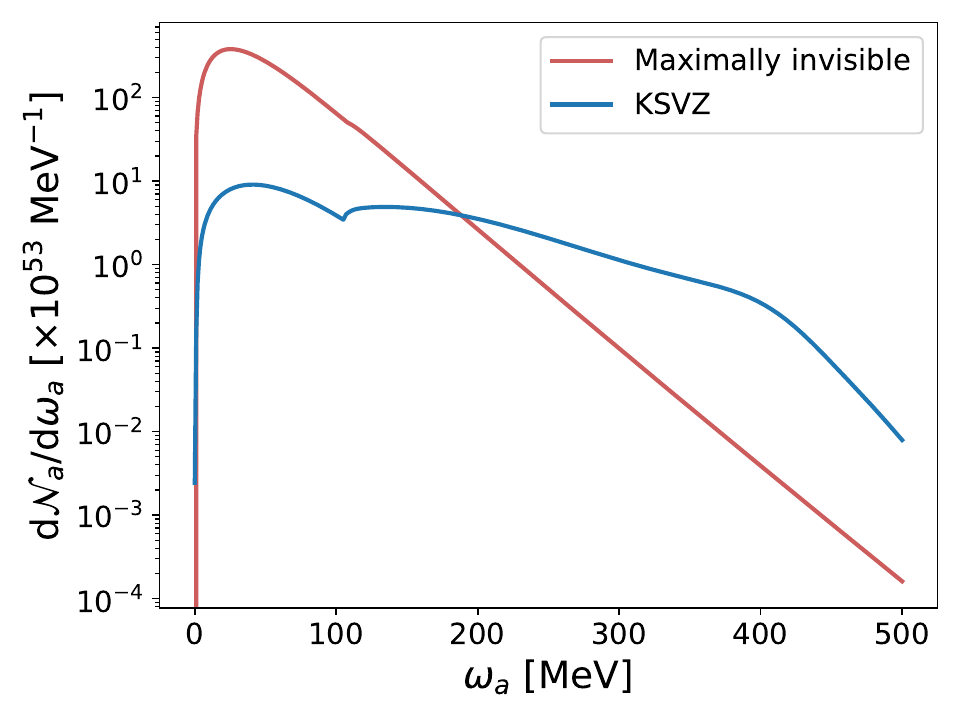}
\caption{Total emission number spectrum for maximally invisible axion production as a function of the energy of the produced axion (\emph{red}) and the analogous spectrum for the KSVZ model (\emph{blue}) as reported in Ref.~\cite{Carenza:2025uib}, both for $f_a$ saturating the corresponding SN 1987A cooling bound.}
\label{fig:c4}
\end{figure}

In Fig.~\ref{fig:c4} we instead display the total maximally invisible axion emission spectrum, and we compare it to the benchmark KSVZ case, as described in Ref.~\cite{Carenza:2025uib}. {In both cases, the chosen value of $f_a$ saturates the corresponding SN 1987A cooling bound, which for the KSVZ model requires $f_a \gtrsim 10^{9} \, \SI{}{GeV}$. We note that in the KSVZ case pion conversion dominates at higher energies. {The difference in the behaviour of the emission spectra can be traced back to the different axion-momentum dependence of the Feynman amplitudes in the maximally invisible scenario and the KSVZ model. Indeed, in the KSVZ model, the axion couples to the nucleons via the interaction terms}} 
 
\begin{equation}
\Lag \supset \frac{\partial_\mu a}{2f_a} \sum_{N = p, n} C_{aNN} \bar{N} \gamma^\mu \gamma_5 N.
\label{eq:c4}
\end{equation}

\begin{figure}[t]
\centering
\subfloat[][]{\includegraphics[width=.45\textwidth]{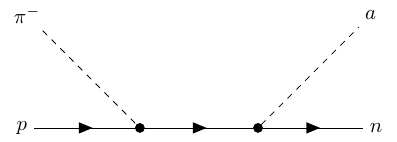}}
\quad
\subfloat[][]{\includegraphics[width=.45\textwidth]{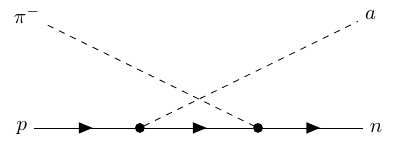}}
\\
\subfloat[][]{\includegraphics[width=.35\textwidth]{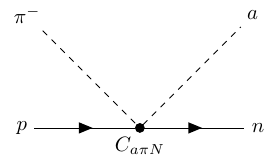}}
\quad
\subfloat[][]{\includegraphics[width=.35\textwidth]{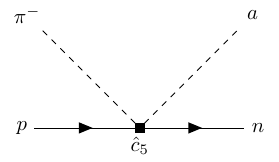}}
\caption{Tree-level Feynman diagrams for axion production through the pion conversion process in the benchmark axion models, (a) to (c), and in maximally invisible axion models, (d).}
\label{fig:c3}
\end{figure}

\noindent In the nonrelativistic limit for the nucleons, this reduces to a derivative coupling of the axion to the nucleon spin, $\sim -\nabla a \cdot \vec{S}$. With reference to diagrams (a) and (b) of Fig.~\ref{fig:c3}, the $aNN$ vertex is therefore proportional to $\abs{\vec{p}_a} \approx \omega_a$, for an ultrarelativistic axion. In the maximally-invisible case, instead, the axion-nucleon interaction emerges from the operator in Eq.~\eqref{eq:b6}, so that the amplitude associated to diagram (d) is independent of $\omega_a$ (see Eq.~\eqref{eq:B6}). {Therefore, processes generated by the $\hat{c}_5$ operator exhibit a suppressed high-energy spectral component, including the pion-conversion contribution, relative to derivative-coupling processes.} {We further observe that, due to this suppression at high energies, the number of produced axions necessary to saturate the SN 1987A bound is larger in the maximally invisible case with respect to the benchmark models.}

\section{Detection strategies}
\label{sec:detection}

In this section we discuss a possible detection approach for the QCD axions produced in SNe. We have seen that, even in the maximally invisible setup, the axion has an irreducible interaction to nucleons. Therefore, this coupling can be also exploited for {axion direct detection} in water Cherenkov detectors, following the strategy outlined in Ref.~\cite{Cavan-Piton:2025nsj}.

{
Axions interacting in water may produce pions through the process, $a + N \to N + \pi^0$, similar to the one in Fig.~\ref{fig:c4}(d), as allowed by the operator defined in Eq.~\eqref{eq:b6}.
Then, the final-state neutral pions can decay $\pi^0 \to \gamma \gamma$ giving a peculiar gamma-ray signature.
}
The two photons would then interact with water, pair-producing charged particles at speeds larger than the speed of light in water, and therefore generating two observable rings of Cherenkov radiation. The unique geometry of this process, with the production of two photon rings in opposite directions, allows one to discriminate the axionic events from the signals due to neutrinos, which instead generate a single Cherenkov ring.

Our goal is then to determine whether such a detection approach would lead to an observable signal in the maximally invisible scenario. The starting point is the determination of the interaction cross section for the process

\begin{equation}
N(p_1) + a(p_2) = N(p_3) + \pi^0(p_4).
\label{eq:d1}
\end{equation}

\noindent The squared Feynman amplitude reads

\begin{equation}
\abs{\overline{\mathcal{M}}_{aN}}^2 = \frac{32 \hat{c}_5^2 m_\pi^4}{f_\pi^2 f_a^2} \frac{z^2}{(1+z)^4} \qty(p_1 \cdot p_3 + m_N^2),
\label{eq:d2}
\end{equation}

\noindent where we have summed over final state spins and averaged over initial state ones. Assuming the axion to be massless, $m_a \approx 0$, we get a simple expression for the derivative of the cross section with respect to the energy of the $\pi^0$ in the rest frame of the initial-state nucleon

\begin{equation}
\qty(\pdv{\sigma}{E_{\pi^0}})_{\text{lab}} = \frac{\hat{c}_5^2 m_\pi^4}{\pi f_\pi^2 f_a^2} \frac{z^2}{(1+z)^4} \frac{\omega_a - E_{\pi^0} + 2 m_N}{\omega_a^2}.
\label{eq:d3}
\end{equation}
Given the cross section and the number spectrum for the production in SNe, the detection spectrum can be then computed as

\begin{equation}
\dv{\mathcal{N}_{\pi^0}^{(a)}}{E_{\pi^0}} = \frac{N_t}{4 \pi d^2} \int_{0}^{+\infty} \dd{\omega_a} \pdv{\sigma}{E_{\pi^0}} \dv{\mathcal{N}_a}{\omega_a},
\label{eq:d4}
\end{equation}

\noindent where $d$ is the distance of the SN from Earth, and $N_t$ is the number of scattering targets. If we take the future Cherenkov detector Hyper-Kamiokande (HK) as a reference, then

\begin{equation}
N_t = 10^9 \times N_{p, n} \times N_A \times \frac{M_{\text{det}}}{\SI{}{kton}} \times \frac{\SI{}{g/mol}}{m_{\text{H}_2\text{O}}},
\end{equation}

\noindent where $N_{p, n}$ is the number of protons (neutrons) per water molecule, $N_A$ is the Avogadro number, $M_{\text{det}} \simeq \SI{258}{kton}$ is the mass of the HK detector, and $m_{\text{H}_2\text{O}} = \SI{18.015}{g/mol}$ is the molar mass of water.

In our calculations we use $d = \SI{0.2}{kpc}$, taking the red supergiant Betelgeuse as a reference future SN. Regarding the number of targets, a precise determination would require a detailed description of the response of nuclear matter in oxygen to the interaction with the axion, which is beyond the scopes of our work. For this reason, in what follows we will work in the two extreme cases in which the axion interacts either with only the protons in the hydrogen nuclei or with all the nucleons in the water molecule. The realistic values of $N_p$ and $N_n$ fall, of course, in between these two extremes, and, as we will see, our choice will not affect our conclusions.

\section{Results and discussion}
\label{sec:results}

\begin{figure}[t]
\centering
\includegraphics[width=.65\textwidth]{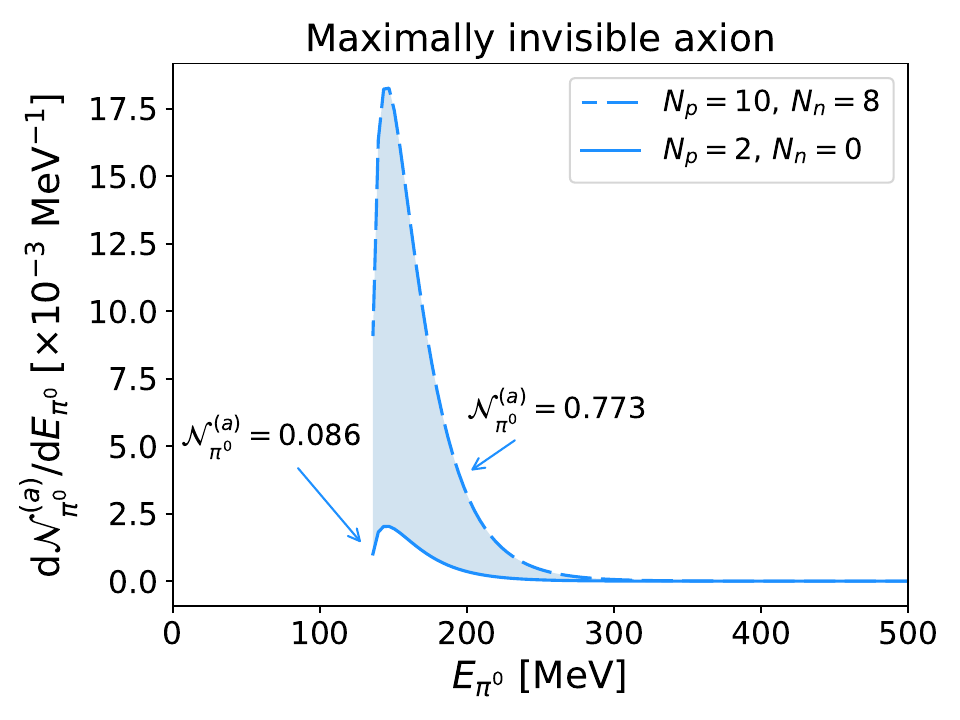}
\caption{Detection number spectra for maximally invisible axions in water Cherenkov detectors as a function of the energy of the produced neutral pion. For the SN profile we used the one in Ref.~\cite{Garch}. The solid curve corresponds to the conservative scenario in which the axion interacts only with the protons in the hydrogen nuclei, while the dashed line corresponds to the other extreme case, in which the axion interacts with all the nucleons in the water molecule. In the labels, we show the predicted total number of events for each case. Here we took Hyper-Kamiokande as a reference detector, and $d = \SI{0.2}{kpc}$. Moreover, $f_a$ is chosen so as to saturate the SN 1987A cooling bound.}
\label{fig:e1}
\end{figure}

{We now apply the detection formalism developed in Sec.~\ref{sec:detection} to the maximally invisible axion scenario. Our benchmark is deliberately optimistic: we consider a very nearby Galactic supernova, represented by Betelgeuse at $d=\SI{0.2}{kpc}$, take Hyper-Kamiokande as reference detector, and choose $f_a$ to saturate the SN~1987A cooling bound. The result is shown in Fig.~\ref{fig:e1}. The two curves correspond to the limiting assumptions discussed above for the number of available targets in water. The realistic prediction depends on the nuclear response of oxygen to the axion-induced pion-production process and is therefore expected to lie between, or at least close to, these two limiting cases.}

{The total number of events is obtained by integrating the differential spectrum over $E_{\pi^0}$. Even under the optimistic assumption that axions scatter incoherently on all nucleons in water, treated as free targets, the predicted signal remains below one event. This is the central result of our analysis: the irreducible nucleon coupling of a maximally invisible QCD axion does provide a guaranteed detection channel in principle, but the corresponding rate is too small to yield an observable Cherenkov signal, even for an exceptionally nearby supernova.}

\section{Conclusions}
\label{sec:conclusions}

The anomalous coupling of the QCD axion to gluons -- essential for the PQ solution to the strong CP problem -- inevitably induces irreducible, model-independent interactions with nucleons. These interactions define a class of \emph{maximally invisible} axion scenarios, in which all model-dependent derivative couplings are suppressed and only the minimal QCD-induced couplings remain~\cite{Springmann:2024ret}.

In this work, we compute the axion emissivity in core-collapse supernovae within this minimal framework.
{In performing this analysis, we revise the tree-level bremsstrahlung amplitudes of Ref.~\cite{Springmann:2024ret}, leading to a correction of the predicted rates by a factor of $\sim 3$.} The corresponding SN~1987A cooling bound on the axion decay constant is
\[
f_a \gtrsim 7.17 \times 10^{7} \, \mathrm{GeV} \, .
\]
{On the detection side, the structure of \emph{maximally invisible} scenarios severely restricts the available experimental handles. In particular, the absence of a photon coupling excludes magnetic-conversion experiments such as helioscopes, as well as searches relying on direct axion-electron or enhanced axion-nucleon couplings. 
The only viable detection channel is the inverse pion-conversion process $a + N \to N + \pi^0$, potentially observable in large water Cherenkov detectors through the characteristic double-ring signature arising from $\pi^0~\to~\gamma\gamma$~\cite{Cavan-Piton:2025nsj}. }

{For the detection process $a + N \to N + \pi^0$, we evaluate the signal using a simulated time-dependent supernova profile. 
{Our results show} that the expected signal at Hyper-Kamiokande~\cite{Hyper-Kamiokande:2018ofw} remains below one event even for an optimistically nearby Galactic supernova benchmark, such as Betelgeuse at $d = 0.2\,\mathrm{kpc}$, assuming $f_a$ saturates the SN~1987A bound. We therefore conclude that \emph{maximally invisible} QCD axions lie beyond the reach of current and foreseeable water Cherenkov detectors.  

The physical message is sobering but clear: QCD axions that solve the strong CP problem can
populate regions of parameter space in which direct detection through SM
couplings is exceedingly difficult, even for values of $f_a$ usually regarded as
accessible. A guaranteed, model-independent detection channel for \emph{every} QCD axion
therefore appears out of reach with current and foreseeable technology. We emphasize,
however, that this is a quantitative statement, conditioned on standard astrophysical
assumptions. The result is sensitive to the modelling of the supernova core -- in
particular to the still-uncertain thermal pion abundance~\cite{Fiorillo:2025gnd,Fore:2023gwv}
and to the effective temperature and duration of the emitting region -- and a hotter or
denser core than the benchmark profile employed here would raise the predicted signal, in
some configurations appreciably.

\acknowledgments

We would like to thank Diego Guadagnoli, Ludovico Vittorio, Konstantin Springmann, Michael Stadlbauer, Stefan Stelzl, and Andreas Weiler for clarifying some points of their works. We warmly thank Thomas Janka for giving us access to the Garching group archive.
This article is based upon work from COST Action COSMIC WISPers
CA21106, supported by COST (European Cooperation 
in Science and Technology). 
The work of LDL is supported by the Italian Ministry of University and Research (MUR) via the FIS2 Consolidator Grant project FIS-2023-02106 -- QAXION (CUP: I53C25001880001). 
MG acknowledges support from project CNS2025-165965, “Señales de axiones en el rango MeV desde fuentes astrofísicas hasta detectores actuales y futuros”, funded by MICIU/AEI/10.13039/501100011033.
In addition, he acknowledges support from 
the Spanish Agencia Estatal de Investigación under grant PID2019-108122GB-C31, and from the ``European Union NextGenerationEU/PRTR'' (Planes Complementarios, Programa de Astrofísica y Física de Altas Energías). 
He also acknowledges support from grant PGC2022-126078NB-C21, ``Aún más allá de los modelos estándar.
Additionally, MG acknowledges funding from the European Union's Horizon 2020 research and innovation programme under the European Research Council (ERC) grant agreement ERC-2017-AdG-788781 (IAXO+).
AL is supported by the
Italian MUR through the FIS 2 project FIS-2023-01577
(DD n. 23314 10-12-2024, CUP C53C24001460001).

\appendix

\section{An explicit model of maximally invisible axions}
\label{app:model}

In this section we want to give an explicit example of a \emph{maximally invisible} axion model, namely a model where the derivative axion couplings to matters and photons are suppressed, $C_{app}\approx C_{ann} \approx C_{a\pi} \approx C_{aee} \approx C_{a\gamma}  \approx 0$. The model we choose is the one presented in Appendix A of \cite{Lucente:2022vuo}, that we summarise here for completeness.

We extend the SM scalar sector with three Higgs doublets $H_{1, 2,3} \sim (1, 2, -1/2)$ and a scalar singlet $\Phi \sim (1, 1, 0)$. The potential of the theory includes the non-Hermitian operators

\begin{equation}
H_3^\dagger H_1 \Phi^2, \quad H_3^\dagger H_2 \Phi^\dagger.
\label{eq:A1}
\end{equation}

\noindent Normalising the PQ charge of the singlet as $X_\Phi = 1$, and imposing orthogonality between the PQ and hypercharge currents, give the following relations between the PQ charges of the Higgs doublets
\begin{subequations}
\begin{gather}
-X_3 + X_1 + 2 = 0, \label{eq:A2a} \\
-X_3 + X_2 - 1 = 0, \label{eq:A2b}\\
X_1 v_1^2 + X_2 v_2^2 + X_3 v_3^2 = 0, \label{eq:A2c}
\end{gather}
\label{eq:A2}
\end{subequations}
\noindent 
where $\ev{H_{1,2,3}} = v_{1,2, 3}$, and $v^2 = v_1^2 + v_2^2 + v_3^2$ is the vacuum expectation value of the SM Higgs field.
In the Yukawa sector we have the operators
\begin{equation}
\begin{array}{llll}
\bar{q}_\alpha u_\beta H_1, & \bar{q}_3 u_3 H_2, & \bar{q}_\alpha u_3 H_1, & \bar{q}_3 u_\beta H_2 \\
\bar{q}_\alpha d_\beta \tilde{H}_2, & \bar{q}_3 d_3 \tilde{H}_1, & \bar{q}_\alpha d_3 \tilde{H}_2, & \bar{q}_3 d_\beta \tilde{H}_1, \\
\bar{\ell}_1 e_1 \tilde{H}_3, & \bar{\ell}_2 e_2 \tilde{H}_1, & \bar{\ell}_3 e_3 \tilde{H}_2, & \dots \, ,
\end{array}
\label{eq:A3}
\end{equation}
where $\tilde{H}_{12} = i \sigma_2 H^*_{1, 2}$, and the greek indices denote the first two quark generations, which are assumed to have the same PQ charges.
Neglecting flavour mixing, the axion couplings to the axial current read
\begin{subequations}
\begin{align}
c_{u, c, t}^0 &= \frac{1}{2N} (X_{u_{1, 2, 3}} - X_{q_{1, 2, 3}}) = \qty{\frac{2}{3} - \frac{X_3}{3}, \,  \frac{2}{3} - \frac{X_3}{3}, \, -\frac{1}{3} - \frac{X_3}{3}}, \label{eq:A4a}\\
c_{d, s, b}^0 &= \frac{1}{2N} (X_{d_{1, 2, 3}} - X_{q_{1, 2, 3}}) = \qty{\frac{1}{3} + \frac{X_3}{3}, \,\frac{1}{3} + \frac{X_3}{3}, \, -\frac{2}{3} + \frac{X_3}{3}}, \label{eq:A4b}\\
c_{e, \mu, \tau}^0 &= \frac{1}{2N} (X_{e_{1, 2, 3}} - X_{\ell_{1, 2, 3}}) = \qty{\frac{X_3}{3}, \, -\frac{2}{3} + \frac{X_3}{3}, \, \frac{1}{3} + \frac{X_3}{3}}, \label{eq:A4c}
\end{align}
\label{eq:A4}
\end{subequations}
where we used the value of the QCD anomaly factor $N$ given by
\begin{equation}
2N = \sum_{i=1}^3 (X_{u_i} + X_{d_i} - 2 X_{q_i}) = 3.
\label{eq:A5}
\end{equation}
The axion couplings to nucleons and pions are related to the axion-quark ones by (see \cite{DiLuzio:2017ogq})
\begin{subequations}
\begin{align}
C_{app} + C_{ann} &= (c_u^0 + c_d^0 - 1) (\Delta u + \Delta d) - \delta_s, \label{eq:A6a} \\
C_{app} - C_{ann} &= \qty(c_u^0 - c_d^0 - \frac{1-z}{1+z})(\Delta u - \Delta d), \label{eq:A6b} \\
C_{a\pi} &= -\frac{1}{3} \qty(c_u^0 - c_d^0 - \frac{1-z}{1+z}).
\end{align}
\label{eq:A6}
\end{subequations}
 where $\delta_s = 0.038(5)c_s^0 + 0.012 (5) c_c^0 + 0.009(2) c_b^0  + 0.0035(4) c_t^0$, $z = m_u/m_d = 0.48(3)$, $\Delta u + \Delta d = 0.521(53)$, and $\Delta u - \Delta d = 1.2723(23)$ \cite{GrillidiCortona:2015jxo}. 

We observe that by construction $c_u^0 + c_d^0 = 1$ in our model so that $C_{app} + C_{ann} \approx 0$. We can also impose $C_{app} - C_{ann}=0$ by setting
\begin{equation}
c_u^0 - c_d^0 = \frac{1}{3} - \frac{2}{3} X_3 = \frac{1-z}{1+z} \approx \frac{1}{3}.
\label{eq:A7}
\end{equation}
This condition, verified for $X_3 \approx 0$ automatically gives $C_{a\pi} \approx 0$ and $C_{aee} = c_e^0 \approx 0$.
Finally, the axion coupling to the photon is given by
\begin{equation}
C_{a\gamma} = \frac{E}{N} - 1.92(4),
\label{eq:A8}
\end{equation}
where $E$ is the QED anomaly coefficient. In our model, quarks and leptons contribute to $E$ as follows 
\begin{equation}
\begin{split}
E_Q &= \sum_{i=1}^3 \qty[3 \qty(\frac{2}{3})^2 (X_{u_i} - X_{q_i}) + 3 \qty(-\frac{1}{3})^2 (X_{u_i} - X_{d_i})] 
= 4 - 3X_3, \\
E_L &= \sum_{i=1}^3 (-1)^2 (X_{e_i} - X_{\ell_i}) = 3X_3 - 1,
\end{split}
\end{equation}
yielding $E/N = 2$ and $C_{a \gamma} = 0.08(4)$, that is compatible with zero at 2$\sigma$.

\section{Axion production from SN at tree-level}
\label{sec:appB}

\subsection{Tree-level matrix elements}

In this section, we present the tree-level  amplitudes for the dominant production channels of maximally invisible axions in supernovae.

\begin{figure}[h]
\centering 
\subfloat[][\label{fig:B1a}]{\includegraphics[width=0.2\textwidth]{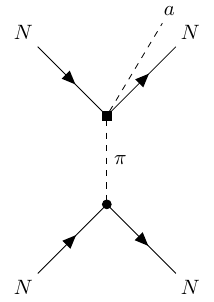}}  \qquad
\subfloat[][\label{fig:B1b}]{\raisebox{1.1cm}{\includegraphics[width=0.3\textwidth]{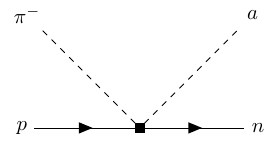}}} 
\caption{Dominant production channels for maximally invisible axions at tree level: axion bremsstrahlung (a) and pion conversion (b).}
\label{fig:B1}
\end{figure}

\paragraph{Axion bremsstrahlung.}

First of all, let us consider the bremsstrahlung process of Fig.~\ref{fig:B1a} 
\[
N(p_1) + N(p_2) \to N(p_3) + N(p_4)  + a(p_a).
\]

\noindent We have two kinds of contributions:

\begin{itemize}

\item Identical nucleons in the initial state: in this case there are 8 diagrams contributing to the amplitude, corresponding to the two possible positions of the axion leg, and the two possible combination of external nucleon legs, each with both the $t-$ and $u-$channel. The total squared amplitude (summed over initial and final spin states), assuming a massless axion ($\abs*{\vec{p}_a} = \omega_a$) and taking the nonrelativistic limit for nucleons, reads
\begin{equation}
\abs{\mathcal{M}_{nn/pp}}^2 = \frac{512 z^2 g_A^2 \hat{c}_5^2 m_\pi^4 m_N^4}{(1+z)^4 f_\pi^4 f_a^2} \qty[\frac{\vec{k}^{\, 2}}{(\vec{k}^{\,2 } + m_\pi^2)^2} + \frac{\vec{\ell}^{\, 2}}{(\vec{\ell}^{\, 2} + m_\pi^2)^2}],
\label{eq:B1}
\end{equation}

\noindent where $g_A$ is the axial vector coupling between pions and nucleons, $\vec{k} = \vec{p}_3 - \vec{p}_1$ and $\vec{\ell} = \vec{p}_4 - \vec{p}_1$. The axion momentum is ignored here because, for non-relativistic nucleons, the momentum $\abs*{\vec{p}_N} = \sqrt{2 m_N E_{\text{kin}}}$ far exceeds the energy scale $\ev{\omega_a} \approx E_{\text{kin}}$.

\item Different nucleons in the initial state: in this case, instead, there are 16 diagrams contributing, consisting of all the possible permutations of the nucleon legs, as well as the two possible position of the axion leg and the $t-$ and $u-$channel contributions. The squared amplitude in the same limits as above is

\begin{equation}
\abs{\mathcal{M}_{np}}^2 = \frac{512 z^2 g_A^2 \hat{c}_5^2 m_\pi^4 m_N^4}{(1+z)^4 f_\pi^4 f_a^2} \qty[\frac{\vec{k}^{\, 2}}{(\vec{k}^{\,2 } + m_\pi^2)^2} + \frac{4\vec{\ell}^{\, 2}}{(\vec{\ell}^{\, 2} + m_\pi^2)^2}].
\label{eq:B2}
\end{equation}

\end{itemize}

{The amplitudes for this process have first been presented in Ref.~\cite{Springmann:2024ret}. We find some differences with respect to the expressions given there;\footnote{{We discussed these differences with the authors of Ref.~\cite{Springmann:2024ret}, who confirmed the validity of our results. Their expressions will be corrected in an upcoming update of their work.}} in particular, apart from an overall factor of 9, in the case with identical initial-state nucleons we do not find a term proportional $\vec{k} \cdot \vec{\ell}$, while in the case with different nucleons in the initial state the second term in Eq.~\eqref{eq:B2} is not present in Ref.~\cite{Springmann:2024ret}.}

\paragraph{Pion conversion.} We now turn our attention to the pion conversion process of Fig.~\ref{fig:B1b}

\[
\pi^-(p_\pi) + p(p_p) \to a(p_a) + n(p_n).
\]

\noindent There is only one diagram contributing to this process at tree level, and the corresponding squared amplitude is

\begin{equation}
\abs{\mathcal{M}_{\pi^- p}}^2 =\frac{128 \hat{c}_5^2 m_\pi^4 m_N^2}{f_\pi^2 f_a^2} \frac{z^2}{(1+z)^4},
\label{eq:B6}
\end{equation}

\noindent which is identical to the one presented in Ref.~\cite{Springmann:2024ret}.

\subsection{Emission number spectra and axion emissivity}

\paragraph{Axion bremsstrahlung.} To find the emission number spectrum per unit time and volume we follow Ref.~\cite{Lucente:2022vuo}. We define

\begin{equation}
\vec{p}_i = \frac{\vec{p}_1 - \vec{p}_2}{2}, \quad \vec{p}_f = \frac{\vec{p}_3 - \vec{p}_4}{2}.
\label{eq:B3}
\end{equation}

\noindent We then write $\abs{\vec{p}_i} = p_i$, $\abs{\vec{p}_f}= p_f$, $\vec{p}_i \cdot \vec{p}_f = p_i p_f z$ and define 

\begin{equation}
u = \frac{p_i^2}{m_N T}, \quad v = \frac{p_f^2}{m_N T}.
\label{eq:B4}
\end{equation}

\noindent One then can find

\begin{equation}
\dv{\dot{n}_a}{\omega_a}\biggl|_{N_i N_j} = S \times \frac{n_i n_j}{512 \pi^{7/2}} \frac{T^{1/2}}{m_N^{5/2}} \omega_a \int_0^{+\infty} \dd{v} \int_{-1}^{1} \dd{z} \exp[-\qty(v + \frac{\omega_a}{T})] \sqrt{v + \frac{\omega_a}{T}} \sqrt{v} \, \abs{\mathcal{M}_{N_i N_j}}^2,
\label{eq:B5}
\end{equation}

\noindent where $n_p = \rho Y_e/m_N$, $n_n = \rho (1-Y_e) /m_N$, and $S$ is a symmetry factor equal to $1/4$ in the case of identical initial-state nucleons and to $1$ otherwise.

\paragraph{Pion conversion.} Following Ref.~\cite{Lella:2022uwi}, we get the following expression for the emission spectrum per unit time and volume:

\begin{equation}
\dv{\dot{n}_a}{\omega_a} \biggl|_{\pi^- p} = \frac{\omega_a (\omega_a^2 - m_\pi^2)^{1/2}}{64 \pi^5 m_N^2}\frac{ (2 m_N T)^{3/2}}{e^{(\omega_a - \mu_\pi)/T}-1} \int_{0}^{+\infty} \dd{y} y^2 \frac{1}{e^{y^2 - \eta_p} + 1} \frac{1}{e^{-y^2 + \eta_n} + 1} \abs*{\mathcal{M}_{\pi^- p}}^2,
\label{eq:B7}
\end{equation}

\noindent where $y = \abs{\vec{p}_p}/(2m_N T)$.

\paragraph{Axion emissivity.} Given the emission number spectra, the emissivity is then obtained as

\begin{equation}
Q_a = \alpha^2(r) \int \dd{\omega_a} \omega_a \dv{\dot{n}_a}{\omega_a},
\label{eq:B8}
\end{equation}

\noindent where $\alpha(r)$ is the lapse function taking into account the redshift between the local SN frame of reference and the distant observer on Earth.

\bibliographystyle{bibi.bst}
\bibliography{bibliography}

@article{Bjorkeroth:2018ipq,
    author = {Bj{\"o}rkeroth, Fredrik and Di Luzio, Luca and Mescia, Federico and Nardi, Enrico},
    title = "{$U(1)$ flavour symmetries as Peccei-Quinn symmetries}",
    eprint = "1811.09637",
    archivePrefix = "arXiv",
    primaryClass = "hep-ph",
    doi = "10.1007/JHEP02(2019)133",
    journal = "JHEP",
    volume = "02",
    pages = "133",
    year = "2019"
}

@article{Badziak:2021apn,
    author = "Badziak, Marcin and Grilli di Cortona, Giovanni and Tabet, Mustafa and Ziegler, Robert",
    title = "{Flavor-violating Higgs decays and stellar cooling anomalies in axion models}",
    eprint = "2107.09708",
    archivePrefix = "arXiv",
    primaryClass = "hep-ph",
    reportNumber = "TTP21-025, P3H-21-051",
    doi = "10.1007/JHEP10(2021)181",
    journal = "JHEP",
    volume = "10",
    pages = "181",
    year = "2021"
}

@article{DiLuzio:2021ysg,
    author = "Di Luzio, Luca and Fedele, Marco and Giannotti, Maurizio and Mescia, Federico and Nardi, Enrico",
    title = "{Stellar evolution confronts axion models}",
    eprint = "2109.10368",
    archivePrefix = "arXiv",
    primaryClass = "hep-ph",
    reportNumber = "DESY-21-141, TTP21-030, P3H-21-062",
    doi = "10.1088/1475-7516/2022/02/035",
    journal = "JCAP",
    volume = "02",
    number = "02",
    pages = "035",
    year = "2022"
}

@article{Badziak:2023fsc,
    author = "Badziak, Marcin and Harigaya, Keisuke",
    title = "{Naturally astrophobic QCD axion}",
    eprint = "2301.09647",
    archivePrefix = "arXiv",
    primaryClass = "hep-ph",
    doi = "10.1007/JHEP06(2023)014",
    journal = "JHEP",
    volume = "06",
    pages = "014",
    year = "2023"
}

@article{Takahashi:2023vhv,
    author = "Takahashi, Fuminobu and Yin, Wen",
    title = "{Hadrophobic axion from a GUT}",
    eprint = "2301.10757",
    archivePrefix = "arXiv",
    primaryClass = "hep-ph",
    reportNumber = "TU-1179",
    doi = "10.1103/PhysRevD.109.035024",
    journal = "Phys. Rev. D",
    volume = "109",
    number = "3",
    pages = "035024",
    year = "2024"
}

@article{Badziak:2024szg,
    author = "Badziak, Marcin and Harigaya, Keisuke and {\L}ukawski, Micha{\l} and Ziegler, Robert",
    title = "{Thermal production of astrophobic axions}",
    eprint = "2403.05621",
    archivePrefix = "arXiv",
    primaryClass = "hep-ph",
    doi = "10.1007/JHEP09(2024)136",
    journal = "JHEP",
    volume = "09",
    pages = "136",
    year = "2024"
}

@article{DiLuzio:2022tyc,
    author = "Di Luzio, Luca and Mescia, Federico and Nardi, Enrico and Okawa, Shohei",
    title = "{Renormalization group effects in astrophobic axion models}",
    eprint = "2205.15326",
    archivePrefix = "arXiv",
    primaryClass = "hep-ph",
    doi = "10.1103/PhysRevD.106.055016",
    journal = "Phys. Rev. D",
    volume = "106",
    number = "5",
    pages = "055016",
    year = "2022"
}

@article{DiLuzio:2024vzg,
    author = "Di Luzio, Luca and Fiorentino, Vincenzo and Giannotti, Maurizio and Mescia, Federico and Nardi, Enrico",
    title = "{Do finite density effects jeopardize axion nucleophobia in supernovae?}",
    eprint = "2410.04613",
    archivePrefix = "arXiv",
    primaryClass = "hep-ph",
    doi = "10.1103/PhysRevD.111.015018",
    journal = "Phys. Rev. D",
    volume = "111",
    number = "1",
    pages = "015018",
    year = "2025"
}

@article{Broggini:2026qxm,
    author = "Broggini, Carlo and Di Carlo, Giuseppe and Di Luzio, Luca and Piatti, Denise and Toni, Claudio",
    title = "{RadioAxion results on the search for axion dark matter under Gran Sasso}",
    eprint = "2602.18392",
    archivePrefix = "arXiv",
    primaryClass = "hep-ex",
    doi = "10.1103/198x-z7c3",
    journal = "Phys. Rev. D",
    volume = "113",
    number = "11",
    pages = "112007",
    year = "2026"
}

@article{Abel:2017rtm,
    author = "Abel, C. and others",
    title = "{Search for Axionlike Dark Matter through Nuclear Spin Precession in Electric and Magnetic Fields}",
    eprint = "1708.06367",
    archivePrefix = "arXiv",
    primaryClass = "hep-ph",
    doi = "10.1103/PhysRevX.7.041034",
    journal = "Phys. Rev. X",
    volume = "7",
    number = "4",
    pages = "041034",
    year = "2017"
}

@article{Roussy:2020ily,
    author = "Roussy, Tanya S. and others",
    title = "{Experimental Constraint on Axionlike Particles over Seven Orders of Magnitude in Mass}",
    eprint = "2006.15787",
    archivePrefix = "arXiv",
    primaryClass = "hep-ph",
    doi = "10.1103/PhysRevLett.126.171301",
    journal = "Phys. Rev. Lett.",
    volume = "126",
    number = "17",
    pages = "171301",
    year = "2021"
}

@article{JEDI:2022hxa,
    author = "Karanth, S. and others",
    collaboration = "JEDI",
    title = "{First Search for Axionlike Particles in a Storage Ring Using a Polarized Deuteron Beam}",
    eprint = "2208.07293",
    archivePrefix = "arXiv",
    primaryClass = "hep-ex",
    doi = "10.1103/PhysRevX.13.031004",
    journal = "Phys. Rev. X",
    volume = "13",
    number = "3",
    pages = "031004",
    year = "2023"
}

@article{Schulthess:2022pbp,
    author = "Schulthess, Ivo and others",
    title = "{New Limit on Axionlike Dark Matter Using Cold Neutrons}",
    eprint = "2204.01454",
    archivePrefix = "arXiv",
    primaryClass = "hep-ex",
    doi = "10.1103/PhysRevLett.129.191801",
    journal = "Phys. Rev. Lett.",
    volume = "129",
    number = "19",
    pages = "191801",
    year = "2022"
}

@article{Zhang:2022ewz,
    author = "Zhang, Xue and Banerjee, Abhishek and Leyser, Mahapan and Perez, Gilad and Schiller, Stephan and Budker, Dmitry and Antypas, Dionysios",
    title = "{Search for Ultralight Dark Matter with Spectroscopy of Radio-Frequency Atomic Transitions}",
    eprint = "2212.04413",
    archivePrefix = "arXiv",
    primaryClass = "physics.atom-ph",
    doi = "10.1103/PhysRevLett.130.251002",
    journal = "Phys. Rev. Lett.",
    volume = "130",
    number = "25",
    pages = "251002",
    year = "2023"
}

@article{Madge:2024aot,
    author = "Madge, Eric and Perez, Gilad and Meir, Ziv",
    title = "{Prospects of nuclear-coupled-dark-matter detection via correlation spectroscopy of I2+ and Ca+}",
    eprint = "2404.00616",
    archivePrefix = "arXiv",
    primaryClass = "physics.atom-ph",
    doi = "10.1103/PhysRevD.110.015008",
    journal = "Phys. Rev. D",
    volume = "110",
    number = "1",
    pages = "015008",
    year = "2024"
}

@article{Zhang:2023lem,
    author = "Zhang, Xin and Houston, Nick and Li, Tianjun",
    title = "{Nuclear decay anomalies as a signature of axion dark matter}",
    eprint = "2303.09865",
    archivePrefix = "arXiv",
    primaryClass = "hep-ph",
    doi = "10.1103/PhysRevD.108.L071101",
    journal = "Phys. Rev. D",
    volume = "108",
    number = "7",
    pages = "L071101",
    year = "2023"
}

@article{Broggini:2024udi,
    author = "Broggini, Carlo and Di Carlo, Giuseppe and Di Luzio, Luca and Toni, Claudio",
    title = "{Alpha radioactivity deep-underground as a probe of axion dark matter}",
    eprint = "2404.18993",
    archivePrefix = "arXiv",
    primaryClass = "hep-ph",
    doi = "10.1016/j.physletb.2024.138836",
    journal = "Phys. Lett. B",
    volume = "855",
    pages = "138836",
    year = "2024"
}

@article{Alda:2024xxa,
    author = "Alda, Jorge and Broggini, Carlo and Di Carlo, Giuseppe and Di Luzio, Luca and Piatti, Denise and Rigolin, Stefano and Toni, Claudio",
    title = "{Weak nuclear decays deep-underground as a probe of axion dark matter}",
    eprint = "2412.20932",
    archivePrefix = "arXiv",
    primaryClass = "hep-ph",
    doi = "10.1103/PhysRevD.111.035022",
    journal = "Phys. Rev. D",
    volume = "111",
    number = "3",
    pages = "035022",
    year = "2025"
}

@article{Cicoli:2026fqp,
    author = "Cicoli, Michele and others",
    title = "{Axions at the meV Crossroads: Theory, Cosmology, Astrophysics, and Experiments}",
    eprint = "2603.18167",
    archivePrefix = "arXiv",
    primaryClass = "hep-ph",
    reportNumber = "LAPTH-006/26, BARI-TH/785-26",
    month = "3",
    year = "2026"
}

@article{tHooft:1976rip,
    author = "'t Hooft, Gerard",
    editor = "Shifman, Mikhail A.",
    title = "{Symmetry Breaking Through Bell-Jackiw Anomalies}",
    reportNumber = "PRINT-76-0254 (HARVARD)",
    doi = "10.1103/PhysRevLett.37.8",
    journal = "Phys. Rev. Lett.",
    volume = "37",
    pages = "8--11",
    year = "1976"
}

@article{Callan:1976je,
    author = "Callan, Jr., Curtis G. and Dashen, R. F. and Gross, David J.",
    editor = "Taylor, J. C.",
    title = "{The Structure of the Gauge Theory Vacuum}",
    reportNumber = "COO-2220-75",
    doi = "10.1016/0370-2693(76)90277-X",
    journal = "Phys. Lett. B",
    volume = "63",
    pages = "334--340",
    year = "1976"
}

@article{Jackiw:1976pf,
    author = "Jackiw, R. and Rebbi, C.",
    editor = "Taylor, J. C.",
    title = "{Vacuum Periodicity in a Yang-Mills Quantum Theory}",
    reportNumber = "MIT-CTP-548",
    doi = "10.1103/PhysRevLett.37.172",
    journal = "Phys. Rev. Lett.",
    volume = "37",
    pages = "172--175",
    year = "1976"
}

@article{Abel:2020pzs,
    author = "Abel, C. and others",
    title = "{Measurement of the Permanent Electric Dipole Moment of the Neutron}",
    eprint = "2001.11966",
    archivePrefix = "arXiv",
    primaryClass = "hep-ex",
    doi = "10.1103/PhysRevLett.124.081803",
    journal = "Phys. Rev. Lett.",
    volume = "124",
    number = "8",
    pages = "081803",
    year = "2020"
}

@article{Peccei:1977hh,
    author = "Peccei, R. D. and Quinn, Helen R.",
    title = "{CP Conservation in the Presence of Instantons}",
    reportNumber = "ITP-568-STANFORD",
    doi = "10.1103/PhysRevLett.38.1440",
    journal = "Phys. Rev. Lett.",
    volume = "38",
    pages = "1440--1443",
    year = "1977"
}

@article{Peccei:1977ur,
    author = "Peccei, R. D. and Quinn, Helen R.",
    title = "{Constraints Imposed by CP Conservation in the Presence of Instantons}",
    reportNumber = "ITP-572-STANFORD",
    doi = "10.1103/PhysRevD.16.1791",
    journal = "Phys. Rev. D",
    volume = "16",
    pages = "1791--1797",
    year = "1977"
}

@article{Weinberg:1977ma,
    author = "Weinberg, Steven",
    title = "{A New Light Boson?}",
    reportNumber = "HUTP-77/A074",
    doi = "10.1103/PhysRevLett.40.223",
    journal = "Phys. Rev. Lett.",
    volume = "40",
    pages = "223--226",
    year = "1978"
}

@article{Wilczek:1977pj,
    author = "Wilczek, Frank",
    title = "{Problem of Strong  $P$  and  $T$  Invariance in the Presence of Instantons}",
    reportNumber = "Print-77-0939 (COLUMBIA)",
    doi = "10.1103/PhysRevLett.40.279",
    journal = "Phys. Rev. Lett.",
    volume = "40",
    pages = "279--282",
    year = "1978"
}

@article{Kim:2008hd,
    author = "Kim, Jihn E. and Carosi, Gianpaolo",
    title = "{Axions and the Strong CP Problem}",
    eprint = "0807.3125",
    archivePrefix = "arXiv",
    primaryClass = "hep-ph",
    doi = "10.1103/RevModPhys.82.557",
    journal = "Rev. Mod. Phys.",
    volume = "82",
    pages = "557--602",
    year = "2010",
    note = "[Erratum: Rev.Mod.Phys. 91, 049902 (2019)]"
}

@article{DiLuzio:2020wdo,
    author = "Di Luzio, Luca and Giannotti, Maurizio and Nardi, Enrico and Visinelli, Luca",
    title = "{The landscape of QCD axion models}",
    eprint = "2003.01100",
    archivePrefix = "arXiv",
    primaryClass = "hep-ph",
    reportNumber = "DESY 20-036, DESY-20-036",
    doi = "10.1016/j.physrep.2020.06.002",
    journal = "Phys. Rept.",
    volume = "870",
    pages = "1--117",
    year = "2020"
}

@article{Giannotti:2022euq,
    author = "Giannotti, Maurizio",
    title = "{Aspects of Axions and ALPs Phenomenology}",
    eprint = "2205.06831",
    archivePrefix = "arXiv",
    primaryClass = "hep-ph",
    doi = "10.1088/1742-6596/2502/1/012003",
    journal = "J. Phys. Conf. Ser.",
    volume = "2502",
    number = "1",
    pages = "012003",
    year = "2023"
}

@article{ADMX:2018gho,
    author = "Du, N. and others",
    collaboration = "ADMX",
    title = "{A Search for Invisible Axion Dark Matter with the Axion Dark Matter Experiment}",
    eprint = "1804.05750",
    archivePrefix = "arXiv",
    primaryClass = "hep-ex",
    reportNumber = "FERMILAB-PUB-18-101-AD-AE",
    doi = "10.1103/PhysRevLett.120.151301",
    journal = "Phys. Rev. Lett.",
    volume = "120",
    number = "15",
    pages = "151301",
    year = "2018"
}

@article{HAYSTAC:2018rwy,
    author = "Zhong, L. and others",
    collaboration = "HAYSTAC",
    title = "{Results from phase 1 of the HAYSTAC microwave cavity axion experiment}",
    eprint = "1803.03690",
    archivePrefix = "arXiv",
    primaryClass = "hep-ex",
    doi = "10.1103/PhysRevD.97.092001",
    journal = "Phys. Rev. D",
    volume = "97",
    number = "9",
    pages = "092001",
    year = "2018"
}

@article{CAPP:2020utb,
    author = "Kwon, Ohjoon and others",
    collaboration = "CAPP",
    title = "{First Results from an Axion Haloscope at CAPP around 10.7  $\mu$eV}",
    eprint = "2012.10764",
    archivePrefix = "arXiv",
    primaryClass = "hep-ex",
    doi = "10.1103/PhysRevLett.126.191802",
    journal = "Phys. Rev. Lett.",
    volume = "126",
    number = "19",
    pages = "191802",
    year = "2021"
}

@article{McAllister:2017lkb,
    author = "McAllister, Ben T. and Flower, Graeme and Kruger, Justin and Ivanov, Eugene N. and Goryachev, Maxim and Bourhill, Jeremy and Tobar, Michael E.",
    collaboration = "ORGAN",
    title = "{The ORGAN Experiment: An axion haloscope above 15 GHz}",
    eprint = "1706.00209",
    archivePrefix = "arXiv",
    primaryClass = "physics.ins-det",
    doi = "10.1016/j.dark.2017.09.010",
    journal = "Phys. Dark Univ.",
    volume = "18",
    pages = "67--72",
    year = "2017"
}

@article{Ouellet:2019tlz,
    author = "Ouellet, Jonathan L. and others",
    title = "{Design and implementation of the ABRACADABRA-10 cm axion dark matter search}",
    eprint = "1901.10652",
    archivePrefix = "arXiv",
    primaryClass = "physics.ins-det",
    doi = "10.1103/PhysRevD.99.052012",
    journal = "Phys. Rev. D",
    volume = "99",
    number = "5",
    pages = "052012",
    year = "2019"
}

@article{Gramolin:2020ict,
    author = "Gramolin, Alexander V. and Aybas, Deniz and Johnson, Dorian and Adam, Janos and Sushkov, Alexander O.",
    title = "{Search for axion-like dark matter with ferromagnets}",
    eprint = "2003.03348",
    archivePrefix = "arXiv",
    primaryClass = "hep-ex",
    doi = "10.1038/s41567-020-1006-6",
    journal = "Nature Phys.",
    volume = "17",
    number = "1",
    pages = "79--84",
    year = "2021"
}

@article{QUAX:2020adt,
    author = "Crescini, N. and others",
    collaboration = "QUAX",
    title = "{Axion search with a quantum-limited ferromagnetic haloscope}",
    eprint = "2001.08940",
    archivePrefix = "arXiv",
    primaryClass = "hep-ex",
    doi = "10.1103/PhysRevLett.124.171801",
    journal = "Phys. Rev. Lett.",
    volume = "124",
    number = "17",
    pages = "171801",
    year = "2020"
}

@article{DMRadio:2022pkf,
    author = "Brouwer, L. and others",
    collaboration = "DMRadio",
    title = "{Projected sensitivity of DMRadio-m3: A search for the QCD axion below 1{\,}{\,}{\ensuremath{\mu}}eV}",
    eprint = "2204.13781",
    archivePrefix = "arXiv",
    primaryClass = "hep-ex",
    doi = "10.1103/PhysRevD.106.103008",
    journal = "Phys. Rev. D",
    volume = "106",
    number = "10",
    pages = "103008",
    year = "2022"
}

@article{CAST:2017uph,
    author = "Anastassopoulos, V. and others",
    collaboration = "CAST",
    title = "{New CAST Limit on the Axion-Photon Interaction}",
    eprint = "1705.02290",
    archivePrefix = "arXiv",
    primaryClass = "hep-ex",
    doi = "10.1038/nphys4109",
    journal = "Nature Phys.",
    volume = "13",
    pages = "584--590",
    year = "2017"
}

@article{IAXO:2019mpb,
    author = "Armengaud, E. and others",
    collaboration = "IAXO",
    title = "{Physics potential of the International Axion Observatory (IAXO)}",
    eprint = "1904.09155",
    archivePrefix = "arXiv",
    primaryClass = "hep-ph",
    doi = "10.1088/1475-7516/2019/06/047",
    journal = "JCAP",
    volume = "06",
    pages = "047",
    year = "2019"
}

@article{IAXO:2024wss,
    author = "Ahyoune, S. and others",
    collaboration = "IAXO",
    title = "{An accurate solar axions ray-tracing response of BabyIAXO}",
    eprint = "2411.13915",
    archivePrefix = "arXiv",
    primaryClass = "hep-ex",
    doi = "10.1007/JHEP02(2025)159",
    journal = "JHEP",
    volume = "02",
    pages = "159",
    year = "2025"
}

@article{IAXO:2020wwp,
    author = "Abeln, A. and others",
    collaboration = "IAXO",
    title = "{Conceptual design of BabyIAXO, the intermediate stage towards the International Axion Observatory}",
    eprint = "2010.12076",
    archivePrefix = "arXiv",
    primaryClass = "physics.ins-det",
    doi = "10.1007/JHEP05(2021)137",
    journal = "JHEP",
    volume = "05",
    pages = "137",
    year = "2021"
}

@article{ALPS:2009des,
    author = "Ehret, Klaus and others",
    collaboration = "ALPS",
    title = "{Resonant laser power build-up in ALPS: A 'Light-shining-through-walls' experiment}",
    eprint = "0905.4159",
    archivePrefix = "arXiv",
    primaryClass = "physics.ins-det",
    reportNumber = "DESY-09-058",
    doi = "10.1016/j.nima.2009.10.102",
    journal = "Nucl. Instrum. Meth. A",
    volume = "612",
    pages = "83--96",
    year = "2009"
}

@article{OSQAR:2015qdv,
    author = "Ballou, R. and others",
    collaboration = "OSQAR",
    title = "{New exclusion limits on scalar and pseudoscalar axionlike particles from light shining through a wall}",
    eprint = "1506.08082",
    archivePrefix = "arXiv",
    primaryClass = "hep-ex",
    doi = "10.1103/PhysRevD.92.092002",
    journal = "Phys. Rev. D",
    volume = "92",
    number = "9",
    pages = "092002",
    year = "2015"
}

@article{ALPSII:2025eri,
    author = "Brotherton, Daniel C. and others",
    collaboration = "ALPS II",
    title = "{Any Light Particle Searches with ALPS II: first science results}",
    eprint = "2512.14110",
    archivePrefix = "arXiv",
    primaryClass = "hep-ex",
    month = "12",
    year = "2025"
}

@article{Carenza:2024ehj,
    author = "Carenza, Pierluca and Giannotti, Maurizio and Isern, Jordi and Mirizzi, Alessandro and Straniero, Oscar",
    title = "{Axion astrophysics}",
    eprint = "2411.02492",
    archivePrefix = "arXiv",
    primaryClass = "hep-ph",
    reportNumber = "BARI-TH/66-24",
    doi = "10.1016/j.physrep.2025.02.002",
    journal = "Phys. Rept.",
    volume = "1117",
    pages = "1--102",
    year = "2025"
}

@article{Caputo:2024oqc,
    author = "Caputo, Andrea and Raffelt, Georg",
    title = "{Astrophysical Axion Bounds: The 2024 Edition}",
    eprint = "2401.13728",
    archivePrefix = "arXiv",
    primaryClass = "hep-ph",
    reportNumber = "MPP-2024-13, CERN-TH-2024-013",
    doi = "10.22323/1.454.0041",
    journal = "PoS",
    volume = "COSMICWISPers",
    pages = "041",
    year = "2024"
}

@article{DiLuzio:2017ogq,
    author = "Di Luzio, Luca and Mescia, Federico and Nardi, Enrico and Panci, Paolo and Ziegler, Robert",
    title = "{Astrophobic Axions}",
    eprint = "1712.04940",
    archivePrefix = "arXiv",
    primaryClass = "hep-ph",
    reportNumber = "IPPP-17-102, CERN-TH-2017-256",
    doi = "10.1103/PhysRevLett.120.261803",
    journal = "Phys. Rev. Lett.",
    volume = "120",
    number = "26",
    pages = "261803",
    year = "2018"
}

@article{Bjorkeroth:2019jtx,
    author = {Bj{\"o}rkeroth, Fredrik and Di Luzio, Luca and Mescia, Federico and Nardi, Enrico and Panci, Paolo and Ziegler, Robert},
    title = "{Axion-electron decoupling in nucleophobic axion models}",
    eprint = "1907.06575",
    archivePrefix = "arXiv",
    primaryClass = "hep-ph",
    reportNumber = "CERN-TH-2019-118, DESY-19-194",
    doi = "10.1103/PhysRevD.101.035027",
    journal = "Phys. Rev. D",
    volume = "101",
    number = "3",
    pages = "035027",
    year = "2020"
}

@article{Graham:2013gfa,
    author = "Graham, Peter W. and Rajendran, Surjeet",
    title = "{New Observables for Direct Detection of Axion Dark Matter}",
    eprint = "1306.6088",
    archivePrefix = "arXiv",
    primaryClass = "hep-ph",
    doi = "10.1103/PhysRevD.88.035023",
    journal = "Phys. Rev. D",
    volume = "88",
    pages = "035023",
    year = "2013"
}

@article{Pospelov:1999mv,
    author = "Pospelov, Maxim and Ritz, Adam",
    title = "{Theta vacua, QCD sum rules, and the neutron electric dipole moment}",
    eprint = "hep-ph/9908508",
    archivePrefix = "arXiv",
    reportNumber = "TPI-MINN-99-34, UMN-TH-1808-99",
    doi = "10.1016/S0550-3213(99)00817-2",
    journal = "Nucl. Phys. B",
    volume = "573",
    pages = "177--200",
    year = "2000"
}

@article{Crewther:1979pi,
    author = "Crewther, R. J. and Di Vecchia, P. and Veneziano, G. and Witten, Edward",
    title = "{Chiral Estimate of the Electric Dipole Moment of the Neutron in Quantum Chromodynamics}",
    reportNumber = "CERN-TH-2735",
    doi = "10.1016/0370-2693(79)90128-X",
    journal = "Phys. Lett. B",
    volume = "88",
    pages = "123",
    year = "1979",
    note = "[Erratum: Phys.Lett.B 91, 487 (1980)]"
}

@article{Springmann:2024mjp,
    author = "Springmann, Konstantin and Stadlbauer, Michael and Stelzl, Stefan and Weiler, Andreas",
    title = "{From supernovae to neutron stars: a systematic approach to axion production at finite density}",
    eprint = "2410.10945",
    archivePrefix = "arXiv",
    primaryClass = "hep-ph",
    reportNumber = "TUM-HEP-1528/24",
    doi = "10.1007/JHEP02(2025)138",
    journal = "JHEP",
    volume = "02",
    pages = "138",
    year = "2025"
}

@article{Springmann:2024ret,
    author = "Springmann, Konstantin and Stadlbauer, Michael and Stelzl, Stefan and Weiler, Andreas",
    title = "{Universal bound on QCD axions from supernovae}",
    eprint = "2410.19902",
    archivePrefix = "arXiv",
    primaryClass = "hep-ph",
    reportNumber = "TUM-HEP-1531/24",
    doi = "10.1103/18t2-1w3b",
    journal = "Phys. Rev. D",
    volume = "112",
    number = "7",
    pages = "075009",
    year = "2025"
}

@article{Lucente:2022vuo,
    author = "Lucente, Giuseppe and Mastrototaro, Leonardo and Carenza, Pierluca and Di Luzio, Luca and Giannotti, Maurizio and Mirizzi, Alessandro",
    title = "{Axion signatures from supernova explosions through the nucleon electric-dipole portal}",
    eprint = "2203.15812",
    archivePrefix = "arXiv",
    primaryClass = "hep-ph",
    doi = "10.1103/PhysRevD.105.123020",
    journal = "Phys. Rev. D",
    volume = "105",
    number = "12",
    pages = "123020",
    year = "2022"
}

@article{Hyper-Kamiokande:2018ofw,
    author = "Abe, K. and others",
    collaboration = "Hyper-Kamiokande",
    title = "{Hyper-Kamiokande Design Report}",
    eprint = "1805.04163",
    archivePrefix = "arXiv",
    primaryClass = "physics.ins-det",
    month = "5",
    year = "2018"
}

@article{Giannotti:2017hny,
    author = "Giannotti, Maurizio and Irastorza, Igor G. and Redondo, Javier and Ringwald, Andreas and Saikawa, Ken'ichi",
    title = "{Stellar Recipes for Axion Hunters}",
    eprint = "1708.02111",
    archivePrefix = "arXiv",
    primaryClass = "hep-ph",
    reportNumber = "DESY-17-116",
    doi = "10.1088/1475-7516/2017/10/010",
    journal = "JCAP",
    volume = "10",
    pages = "010",
    year = "2017"
}

@article{Scherer:2002tk,
    author = "Scherer, Stefan",
    editor = "Negele, John W. and Vogt, E. W.",
    title = "{Introduction to chiral perturbation theory}",
    eprint = "hep-ph/0210398",
    archivePrefix = "arXiv",
    reportNumber = "MKPH-T-02-09",
    journal = "Adv. Nucl. Phys.",
    volume = "27",
    pages = "277",
    year = "2003"
}

@article{Janka:2006fh,
    author = "Janka, Hans-Thomas and Langanke, K. and Marek, A. and Martinez-Pinedo, G. and Mueller, B.",
    title = "{Theory of Core-Collapse Supernovae}",
    eprint = "astro-ph/0612072",
    archivePrefix = "arXiv",
    doi = "10.1016/j.physrep.2007.02.002",
    journal = "Phys. Rept.",
    volume = "442",
    pages = "38--74",
    year = "2007"
}

@article{Mirizzi:2015eza,
    author = "Mirizzi, Alessandro and Tamborra, Irene and Janka, Hans-Thomas and Saviano, Ninetta and Scholberg, Kate and Bollig, Robert and Hudepohl, Lorenz and Chakraborty, Sovan",
    title = "{Supernova Neutrinos: Production, Oscillations and Detection}",
    eprint = "1508.00785",
    archivePrefix = "arXiv",
    primaryClass = "astro-ph.HE",
    doi = "10.1393/ncr/i2016-10120-8",
    journal = "Riv. Nuovo Cim.",
    volume = "39",
    number = "1-2",
    pages = "1--112",
    year = "2016"
}

@article{Carenza:2019pxu,
    author = "Carenza, Pierluca and Fischer, Tobias and Giannotti, Maurizio and Guo, Gang and Mart{\'\i}nez-Pinedo, Gabriel and Mirizzi, Alessandro",
    title = "{Improved axion emissivity from a supernova via nucleon-nucleon bremsstrahlung}",
    eprint = "1906.11844",
    archivePrefix = "arXiv",
    primaryClass = "hep-ph",
    doi = "10.1088/1475-7516/2019/10/016",
    journal = "JCAP",
    volume = "10",
    number = "10",
    pages = "016",
    year = "2019",
    note = "[Erratum: JCAP 05, E01 (2020)]"
}

@article{Stoica:2009zh,
    author = "Stoica, S. and Pastrav, B. and Horvath, J. E. and Allen, M. P.",
    title = "{Pion mass effects on axion emission from neutron stars through NN bremsstrahlung processes}",
    eprint = "0906.3134",
    archivePrefix = "arXiv",
    primaryClass = "nucl-th",
    doi = "10.1016/j.nuclphysa.2009.07.007",
    journal = "Nucl. Phys. A",
    volume = "828",
    pages = "439--449",
    year = "2009",
    note = "[Erratum: Nucl.Phys.A 832, 148 (2010)]"
}

@article{Ericson:1988wr,
    author = "Ericson, Torleif Erik Oskar and Mathiot, J. F.",
    title = "{Axion Emission from SN 1987a: Nuclear Physics Constraints}",
    reportNumber = "CERN-TH-5227/88, IPNO/TH-88-63",
    doi = "10.1016/0370-2693(89)91103-9",
    journal = "Phys. Lett. B",
    volume = "219",
    pages = "507--514",
    year = "1989"
}

@article{Raffelt:1991pw,
    author = "Raffelt, Georg and Seckel, David",
    title = "{Multiple scattering suppression of the bremsstrahlung emission of neutrinos and axions in supernovae}",
    reportNumber = "MPI-PH-91-31-REV, BA-91-41-REV",
    doi = "10.1103/PhysRevLett.67.2605",
    journal = "Phys. Rev. Lett.",
    volume = "67",
    pages = "2605--2608",
    year = "1991"
}

@article{Janka:1995ir,
    author = "Janka, Hans-Thomas and Keil, Wolfgang and Raffelt, Georg and Seckel, David",
    title = "{Nucleon spin fluctuations and the supernova emission of neutrinos and axions}",
    eprint = "astro-ph/9507023",
    archivePrefix = "arXiv",
    reportNumber = "MPI-PHT-95-63, SFB-375-18, BA-95-24",
    doi = "10.1103/PhysRevLett.76.2621",
    journal = "Phys. Rev. Lett.",
    volume = "76",
    pages = "2621--2624",
    year = "1996"
}

@article{Fore:2019wib,
    author = "Fore, Bryce and Reddy, Sanjay",
    title = "{Pions in hot dense matter and their astrophysical implications}",
    eprint = "1911.02632",
    archivePrefix = "arXiv",
    primaryClass = "astro-ph.HE",
    reportNumber = "INT-PUB-19-046",
    doi = "10.1103/PhysRevC.101.035809",
    journal = "Phys. Rev. C",
    volume = "101",
    number = "3",
    pages = "035809",
    year = "2020"
}

@article{Carenza:2020cis,
    author = "Carenza, Pierluca and Fore, Bryce and Giannotti, Maurizio and Mirizzi, Alessandro and Reddy, Sanjay",
    title = "{Enhanced Supernova Axion Emission and its Implications}",
    eprint = "2010.02943",
    archivePrefix = "arXiv",
    primaryClass = "hep-ph",
    reportNumber = "INT-PUB-20-039",
    doi = "10.1103/PhysRevLett.126.071102",
    journal = "Phys. Rev. Lett.",
    volume = "126",
    number = "7",
    pages = "071102",
    year = "2021"
}

@article{Choi:2021ign,
    author = "Choi, Kiwoon and Kim, Hee Jung and Seong, Hyeonseok and Shin, Chang Sub",
    title = "{Axion emission from supernova with axion-pion-nucleon contact interaction}",
    eprint = "2110.01972",
    archivePrefix = "arXiv",
    primaryClass = "hep-ph",
    reportNumber = "CTPU-PTC-21-35",
    doi = "10.1007/JHEP02(2022)143",
    journal = "JHEP",
    volume = "02",
    pages = "143",
    year = "2022"
}

@article{Ho:2022oaw,
    author = "Ho, Shu-Yu and Kim, Jongkuk and Ko, Pyungwon and Park, Jae-hyeon",
    title = "{Supernova axion emissivity with {\ensuremath{\Delta}}(1232) resonance in heavy baryon chiral perturbation theory}",
    eprint = "2212.01155",
    archivePrefix = "arXiv",
    primaryClass = "hep-ph",
    reportNumber = "KIAS-P22082",
    doi = "10.1103/PhysRevD.107.075002",
    journal = "Phys. Rev. D",
    volume = "107",
    number = "7",
    pages = "075002",
    year = "2023"
}

@article{Carenza:2023lci,
    author = "Carenza, Pierluca",
    title = "{Axion emission from supernovae: a cheatsheet}",
    eprint = "2309.14798",
    archivePrefix = "arXiv",
    primaryClass = "hep-ph",
    doi = "10.1140/epjp/s13360-023-04484-2",
    journal = "Eur. Phys. J. Plus",
    volume = "138",
    number = "9",
    pages = "836",
    year = "2023"
}

@article{Lella:2022uwi,
    author = "Lella, Alessandro and Carenza, Pierluca and Lucente, Giuseppe and Giannotti, Maurizio and Mirizzi, Alessandro",
    title = "{Protoneutron stars as cosmic factories for massive axionlike particles}",
    eprint = "2211.13760",
    archivePrefix = "arXiv",
    primaryClass = "hep-ph",
    doi = "10.1103/PhysRevD.107.103017",
    journal = "Phys. Rev. D",
    volume = "107",
    number = "10",
    pages = "103017",
    year = "2023"
}

@misc{Garch,
  title = {Garching core-collapse supernova research archive},
  howpublished = {\url{https://wwwmpa.mpa-garching.mpg.de/ccsnarchive//}}
}

@article{Carenza:2025uib,
    author = "Carenza, P. and others",
    title = "{Detecting supernova axions with IAXO}",
    eprint = "2502.19476",
    archivePrefix = "arXiv",
    primaryClass = "hep-ph",
    reportNumber = "BARI-TH/778-25",
    doi = "10.1088/1475-7516/2025/07/075",
    journal = "JCAP",
    volume = "07",
    pages = "075",
    year = "2025"
}

@article{Lella:2024dmx,
    author = "Lella, Alessandro and Ravensburg, Eike and Carenza, Pierluca and Marsh, M. C. David",
    title = "{Supernova limits on QCD axionlike particles}",
    eprint = "2405.00153",
    archivePrefix = "arXiv",
    primaryClass = "hep-ph",
    doi = "10.1103/PhysRevD.110.043019",
    journal = "Phys. Rev. D",
    volume = "110",
    number = "4",
    pages = "043019",
    year = "2024"
}

@article{Lella:2024hfk,
    author = "Lella, Alessandro and Calore, Francesca and Carenza, Pierluca and Eckner, Christopher and Giannotti, Maurizio and Lucente, Giuseppe and Mirizzi, Alessandro",
    title = "{Probing protoneutron stars with gamma-ray axionscopes}",
    eprint = "2405.02395",
    archivePrefix = "arXiv",
    primaryClass = "hep-ph",
    reportNumber = "LAPTH-024/24, BARI-TH/769-24",
    doi = "10.1088/1475-7516/2024/11/009",
    journal = "JCAP",
    volume = "11",
    pages = "009",
    year = "2024"
}

@article{Cavan-Piton:2025nsj,
    author = "Cavan-Piton, Mael and Guadagnoli, Diego and Iohner, Axel and Fernandez-Menendez, Pablo and Vittorio, Ludovico",
    title = "{Probing the general axion-nucleon interaction in water Cherenkov experiments}",
    eprint = "2503.17490",
    archivePrefix = "arXiv",
    primaryClass = "hep-ph",
    reportNumber = "LAPTH-011/25",
    doi = "10.1007/JHEP07(2025)070",
    journal = "JHEP",
    volume = "07",
    pages = "070",
    year = "2025"
}

@article{Fore:2023gwv,
    author = "Fore, Bryce and Kaiser, Norbert and Reddy, Sanjay and Warrington, Neill C.",
    title = "{Mass of charged pions in neutron-star matter}",
    eprint = "2301.07226",
    archivePrefix = "arXiv",
    primaryClass = "nucl-th",
    doi = "10.1103/PhysRevC.110.025803",
    journal = "Phys. Rev. C",
    volume = "110",
    number = "2",
    pages = "025803",
    year = "2024"
}

@article{Bollig:2020xdr,
    author = "Bollig, Robert and DeRocco, William and Graham, Peter W. and Janka, Hans-Thomas",
    title = "{Muons in Supernovae: Implications for the Axion-Muon Coupling}",
    eprint = "2005.07141",
    archivePrefix = "arXiv",
    primaryClass = "hep-ph",
    doi = "10.1103/PhysRevLett.125.051104",
    journal = "Phys. Rev. Lett.",
    volume = "125",
    number = "5",
    pages = "051104",
    year = "2020",
    note = "[Erratum: Phys.Rev.Lett. 126, 189901 (2021)]"
}

@article{Caputo:2021rux,
    author = "Caputo, Andrea and Raffelt, Georg and Vitagliano, Edoardo",
    title = "{Muonic boson limits: Supernova redux}",
    eprint = "2109.03244",
    archivePrefix = "arXiv",
    primaryClass = "hep-ph",
    reportNumber = "MPP-2021-154",
    doi = "10.1103/PhysRevD.105.035022",
    journal = "Phys. Rev. D",
    volume = "105",
    number = "3",
    pages = "035022",
    year = "2022"
}

@article{Caputo:2022mah,
    author = "Caputo, Andrea and Janka, Hans-Thomas and Raffelt, Georg and Vitagliano, Edoardo",
    title = "{Low-Energy Supernovae Severely Constrain Radiative Particle Decays}",
    eprint = "2201.09890",
    archivePrefix = "arXiv",
    primaryClass = "astro-ph.HE",
    doi = "10.1103/PhysRevLett.128.221103",
    journal = "Phys. Rev. Lett.",
    volume = "128",
    number = "22",
    pages = "221103",
    year = "2022"
}

@article{Lella:2023bfb,
    author = "Lella, Alessandro and Carenza, Pierluca and Co', Giampaolo and Lucente, Giuseppe and Giannotti, Maurizio and Mirizzi, Alessandro and Rauscher, Thomas",
    title = "{Getting the most on supernova axions}",
    eprint = "2306.01048",
    archivePrefix = "arXiv",
    primaryClass = "hep-ph",
    doi = "10.1103/PhysRevD.109.023001",
    journal = "Phys. Rev. D",
    volume = "109",
    number = "2",
    pages = "023001",
    year = "2024"
}

@article{Raffelt:1987yt,
    author = "Raffelt, Georg and Seckel, David",
    title = "{Bounds on Exotic Particle Interactions from SN 1987a}",
    reportNumber = "SCIPP-87/107",
    doi = "10.1103/PhysRevLett.60.1793",
    journal = "Phys. Rev. Lett.",
    volume = "60",
    pages = "1793",
    year = "1988"
}

@article{GrillidiCortona:2015jxo,
    author = "Grilli di Cortona, Giovanni and Hardy, Edward and Pardo Vega, Javier and Villadoro, Giovanni",
    title = "{The QCD axion, precisely}",
    eprint = "1511.02867",
    archivePrefix = "arXiv",
    primaryClass = "hep-ph",
    doi = "10.1007/JHEP01(2016)034",
    journal = "JHEP",
    volume = "01",
    pages = "034",
    year = "2016"
}

@article{Fiorillo:2025gnd,
    author = "Fiorillo, Damiano F. G. and Gil Muyor, {\'A}ngel and Janka, Hans-Thomas and Raffelt, Georg G. and Vitagliano, Edoardo",
    title = "{Axion-photon conversion in transient compact stars: Systematics, constraints, and opportunities}",
    eprint = "2509.13322",
    archivePrefix = "arXiv",
    primaryClass = "hep-ph",
    doi = "10.1088/1475-7516/2026/03/053",
    journal = "JCAP",
    volume = "03",
    pages = "053",
    year = "2026"
}

\end{document}